\documentclass[a4paper,11pt]{amsart}
\begin{document}
\hyphenation{ri-go-rously gra-vi-ta-tio-nal ope-ra-tors
re-pre-sented no-thing}
\title[] {{\bf  On the origin of the notion of GW \\ \emph{et cetera}}}
\author[]{Angelo Loinger}
\thanks{To be
published on \emph{Spacetime \& Substance.} \\email:
angelo.loinger@mi.infn.it\\ Dipartimento di Fisica, Universit\`a
di Milano, Via Celoria, 16 - 20133 Milano (Italy)}

\begin{abstract}
The notion of gravitational wave (GW) came forth originally as a
by-product of the \emph{\textbf{linear}} approximation of general
relativity (GR). Now, it can be proved that this approximation is
quite \emph{\textbf{inadequate}} to a proper study of the
hypothetic GW's. The significant role of the approximations beyond
the linear stage is emphasized.
\end{abstract}

\maketitle

\vskip1.20cm
\textbf{1}. -  As it is well known, the \emph{linear}
approximation of GR (physically, the approximation for \emph{weak}
gravitational fields) has Minkowski spacetime as its
(\emph{fixed}) substrate \cite{1}, \cite{2}. It resembles the e.m.
Maxwell theory, and is only Lorentz invariant.

\par With respect to transformations of \emph{general}
co-ordinates, its energy tensor becomes a \emph{false} (pseudo)
tensor, which can be reduced to zero through a suitable change of
reference system.

\par A celebrated by-product of the linearized version of GR is
the notion of gravitational wave \cite{1}, \cite{2}. Now, in 1944
Weyl \cite{3} pointed out that, rigorously speaking, the
gravitational field of the linearized version exerts \emph{no}
force on matter, i.e. is a ``powerless shadow''. Indeed, a basic
result of the Einstein-Infeld-Hoffmann method \cite{4} tells us
that, as Weyl \cite{3} wrote, ``the gravitational force arises
only when one continues the approximation beyond the linear
stage.'' Even in the modern literature this fact is generally
overlooked and, quite uncritically, the action on matter of a
gravitational wave -- e.g. of a plane wave -- is formally
computed.

\par In conclusion,the linear approximation of GR -- which is the
favourite relativistic doctrine of the GW hunters -- is completely
\emph{inadequate} to an approximate treatment of the question of
the GW's.

\par On the other hand, if we continue the approximation beyond the
linear stage (cf. \cite{4} and \cite{5}), we find that the
radiation terms of the gravitational field can be \emph{destroyed}
by convenient co-ordinate transformations: this proves that the
GW's are \emph{only a product of a special choice of the reference
frame}, i.e. that they do not possess a \emph{physical} reality
\cite{6}, \cite{6bis}.

\par In the recent literature the above crucial role of the
co-ordinate system is ignored. E.g., Itoh and Futamase have
published a learned study on the third post-Newtonian equation of
motion for relativistic compact binaries \cite{7}, \cite{8};
\emph{their result is derived under the \textbf{harmonic}
co-ordinate condition}. Motivation for this research \cite{7}:
``One promising source of gravitational waves for those detectors
$[$i.e. GEO600, LIGO, TAMA300$]$ is a relativistic compact binary
system in an inspiraling phase. The detectability and quality of
measurements of astrophysical information of such gravitational
wave sources rely on the accuracy of our theoretical knowledge
about the waveforms.  A high order, say, third- or fourth-order,
post-Newtonian equation of motion for an inspiraling compact
binary is one of the necessary ingredients to construct and study
such waveforms $[\ldots]$.'' Itoh writes (see the abstract of
\cite{8}): ``Our resulting equation of motion admits a conserved
energy (neglecting the 2.5 PN radiation reaction effect), is
Lorentz invariant, and is unambiguous $[\ldots]$.''

\par These authors do not suspect that the radiation terms are
\emph{frame dependent}, and can be destroyed by a suitable change
of co-ordinate system. Further, they are unaware that -- as it is
easy to prove -- the motions of point masses interacting only
gravitationally -- as, e.g., the two compact stars of some
binaries or the bodies of the solar system -- happen along
\emph{\textbf{geodesic}} lines \cite{6}, and consequently an
emission of GW's is obviously impossible.

\par More radically, it can be proved that \emph{no}
``mechanism'' exists \emph{in GR} for the generation of GW's
\cite{6}.

\vskip0.50cm
\textbf{2}. -  Only recently, through a kind letter of Prof. A.
Gsponer, I have known the existence of the beautiful memoir by
Weyl \cite{3}.

\par Since it seems that the astrophysical community is not aware of Weyl's
results, I think very useful to reproduce in an
\emph{\textbf{APPENDIX}}, at the end of the present paper, the
``Introduction and Summary'' and sects. \textbf{1}, \textbf{2} of
Weyl's memoir, which are particularly relevant to our theme.

\par I avail myself of this opportunity for an apology: in 1999
I published a very short Note entitled ``Deduction of the law of
motion of the charges from Maxwell equations'' \cite{9}. Now, my
result is contained in Weyl's treatment of Maxwell theory, see
sect. \textbf{2}. of \cite{3}. Weyl wrote that this theorem was
``well known''. Yes, but only to the Blessed Few!

\vskip0.70cm
\textbf{Acknowledgment}. -  I am very grateful to Prof. G.
Morpurgo, who has called my attention to the research of Itoh and
Futamase \cite{7}, \cite{8}.

\small \vskip0.5cm\par\hfill {\emph{``Nil sapientiae odiosius
acumine nimio''.}
     \vskip0.10cm\par\hfill Seneca}

\vskip2.00cm
\begin{center}
\noindent \emph{\textbf{APPENDIX}}\nopagebreak \par \nopagebreak
 \vskip0.50cm\nopagebreak

\textbf{HOW FAR CAN ONE GET WITH A LINEAR FIELD THEORY OF
GRAVITATION IN FLAT SPACE-TIME?}\footnote{Received August 9,
1944.} \vskip0.30cm By HERMANN WEYL
\\(from: \emph{Amer. J. Math.} \textbf{66} (1944) 591)
\end{center}
\vskip0.30cm

\textbf{Introduction and Summary.} G.D. Birkhoff's attempt to
establish a linear field theory of gravitation within the frame of
special relativity\footnote{\emph{Proceedings of the National
Academy of Sciences}, vol.\textbf{29} (1943), p.231.} makes it
desirable to probe the potentialities and limitations of such a
theory in more general terms. In thus continuing a discussion
begun at another place\footnote{\emph{Proceedings of the National
Academy of Sciences}, vol.\textbf{30} (1944), p.205.} I find that
the differential operators at one's disposal form a 5 dimensional
linear manifold. But the requirement that the field equations
imply the law of conservation of energy and momentum in the simple
form $\partial T_{i}^{k}/\partial x_{k}=0$ limit these
$\infty^{5}$ possibilities to $\infty^{2}$, which, however, reduce
easily to two cases, a regular one $(L)$ and a singular one
$(L')$. The regular case $(L)$ is nothing but Einstein's theory of
weak fields. Resembling very closely Maxwell's theory of the
electromagnetic field, it satisfies a principle of gauge
invariance involving 4 arbitrary functions, and although its
gravitational field exerts no force on matter, it is well suited
to illustrate the role of energy and momentum, charge and mass in
the interplay between matter and field. It might also help, though
this is much more problematic, in pointing the way to a more
satisfactory unification of gravitation and electricity than we at
present possess. Birkhoff follows the opposite way: by avoiding
rather than adopting the $\infty^{2}$ special operators mentioned
above, his ``dualistic'' theory $(B)$ destroys the bond between
mechanical and field equations, which is such a decisive feature
in Einstein's theory.

\vskip0.50cm
\textbf{1. Maxwell's theory of the electromagnetic
field and the monistic linear theory of gravitation (\emph{L}).
Gauge invariance}. Within the frame of special relativity and its
metric ground form

\begin{equation} \nonumber
    \textrm{d}s^{2} = \delta_{ik} \textrm{d}x_{i} \textrm{d}x_{k} = \textrm{d}x_{0}^{2} -
    (\textrm{d}x_{1}^{2} + \textrm{d}x_{2}^{2} + \textrm{d}x_{3}^{2})
\end{equation}

an electromagnetic field is described by a skew tensor

\begin{equation} \nonumber
    f_{ik} = \partial \phi_{k} / \partial x_{i} - \partial \phi_{i} /
\partial x_{k}
\end{equation}

derived from a vector potential $\phi_{i}$ and satisfies Maxwell's
equations

\begin{equation} \label{eq:Wone}
    \partial f^{ki} / \partial x_{k} = s^{i} \qquad \textrm{or} \qquad
    D_{i}\phi = \Box \phi_{i} - \partial \phi' / \partial x_{i} = s_{i}
\end{equation}

where $s^{i}$ is the density-flow of electric charge and

\begin{equation} \nonumber
    \phi' = \partial \phi^{i} / \partial x_{i}, \qquad
    \Box \phi = \delta^{pq}(\partial ^{2} \phi / \partial x_{p} \partial x_{q})   .
\end{equation}

The equations do not change if one substitutes

\begin{equation} \label{eq:Wtwo}
    \phi^{*}_{i} = \phi_{i} - \partial \lambda / \partial x_{i}
    \qquad \textrm{for } \phi_{i}  ,
\end{equation}

$\lambda$ being an arbitrary function of the
co$\ddot{\textrm{o}}$rdinates (``\emph{gauge invariance}''), and
they imply the differential conservation law of electric charge:

\begin{equation} \label{eq:Wthree}
    \partial s^{i} / \partial x_{i}=0.
\end{equation}

\par As is easily  verified, there are only two ways in which one
may form a vector field by linear combination of the second
derivatives of a given vector field $\phi_{i}$, namely,

\begin{equation} \nonumber
    \Box \phi_{i} \quad \textrm{and} \quad \partial \phi'_{i} / \partial x_{i}
    \qquad (\phi'_{i} = \partial \phi^{p} / \partial x_{p}) .
\end{equation}

The only linear combination $D_{i}\phi$ of these two vector fields
which satisfies the identity $(\partial /
\partial x_{i})(D^{i}\phi)=0$ is the one occurring in (\ref{eq:Wone}),

\begin{equation} \nonumber
   D_{i}\phi = \Box \phi_{i} - \partial \phi' / \partial x_{i}  .
\end{equation}

Herein lies as sort of mathematical justification for Maxwell's
equations.
\par Taking from Einstein's theory of gravitation the hint that
gravitation is represented by a symmetric tensor potential
$h_{ik}$, but trying to emulate the linear character of Maxwell's
theory of the electromagnetic field, one could ask oneself what
symmetric tensors $\bar{D}_{ik}h$ can be constructed by linear
combination from the second derivatives of $h_{ik}$. The answer is
that there are 5 such expressions, namely

\begin{equation} \label{eq:Wfour}
   \Box h_{ik}, \quad \partial h'_{i} / \partial x_{k} + \partial h'_{k} / \partial
   x_{i}, \quad
  h''\delta_{ik}, \quad  \partial^{2}h / \partial x_{i} \partial
  x_{k},\quad
  \Box h \cdot \delta_{ik}
\end{equation}

where

\begin{equation} \nonumber
   h=h_{p}^{p}, \qquad h'_{i} = \partial h_{i}^{p} / \partial
   x_{p},\qquad
   h'' = \partial ^{2}h^{pq} / \partial x_{p} \partial x_{q}.
\end{equation}

With any linear combination $\bar{D}_{ik}h$ of these 5 expressions
one could set up the field equations of gravitation

\begin{equation} \label{eq:Wfive}
   \bar{D}_{ik}h = T_{ik}
\end{equation}

the right member of which is the energy-momentum tensor $T_{ik}$.
In analogy to the situation encountered in Maxwell's theory one
may ask further for which linear combinations $\bar{D}_{ik}$ the
identity

\begin{equation} \nonumber
   (\partial / \partial x_{k})(\bar{D}_{i}^{k}h) = 0
\end{equation}

will hold, and one finds that this is the case if, and only if,
$\bar{D}_{ik}h$ is of the form

\begin{equation} \label{eq:Wsix}
\alpha \{ \Box h_{ik} - (\partial h'_{i} /
\partial x_{k} + \partial h'_{k} / \partial x_{i}) + h''\delta_{ik}\} +
\beta \{ \partial ^{2}h /
\partial x_{i} \partial x_{k} - \Box h \cdot \delta_{ik}\} .
\end{equation}

$\alpha$ and $\beta$ being arbitrary constants. In this case the
field equations (\ref{eq:Wfive}) entail the differential
conservation law of energy and momentum

\begin{equation} \label{eq:Wseven}
   \partial T_{i}^{k} / \partial x_{k}=0.
\end{equation}

With two constants $a$, $b$, ($a\neq0$, $a\neq4b$) we can make the
substitution

\begin{equation} \nonumber
   h_{ik} \rightarrow a \cdot h_{ik} - b \cdot h \delta_{ik}
\end{equation}

and thereby reduce $\alpha$, $\beta$ to the values 1, 1, provided
$\alpha \neq 0$, $\alpha \neq 2\beta$. Hence, disregarding these
singular values, we may assume as our field equations

\begin{equation} \label{eq:Wfivex}
D_{ik}h \equiv \{ \Box h_{ik} - (\partial h'_{i} /
\partial x_{k} + \partial h'_{k} / \partial x_{i}) + h''\delta_{ik}\} +
\{ \partial ^{2}h /
\partial x_{i} \partial x_{k} - \Box h \cdot \delta_{ik}\} = T_{ik} .
\end{equation}

$D_{ik}h$ remains unchanged if $h_{ik}$ is replaced by

\begin{equation} \label{eq:Weight}
h^{*}_{ik} = h_{ik} + (\partial \xi_{i} /
\partial x_{k} + \partial \xi_{k} / \partial x_{i})
\end{equation}

where $\xi_{i}$ is an arbitrary vector field. Hence we have the
same type of correlation between gauge invariance and conservation
law for the gravitational field as for the electromagnetic field,
and it is reasonable to consider as physically equivalent any two
tensor fields $h$, $h^{*}$ which are related by (\ref{eq:Weight}).

\par The linear theory of gravitation $(L)$ in a flat world at
which one thus arrives with a certain mathematical necessity is
nothing else but \emph{Einstein's theory for weak fields}. Indeed,
on replacing Einstein's $g_{ik}$ by $\delta_{ik} + 2\kappa \cdot
h_{ik}$ and neglecting higher powers of the gravitational constant
$\kappa$, one obtains (\ref{eq:Wfivex}), and the property of gauge
invariance (\ref{eq:Weight}) reflects the invariance of Einstein's
equations with respect to arbitrary co$\ddot{\textrm{o}}$rdinate
transformations\footnote{Cf. A. Einstein, \emph{Sitzungsber.
Preuss. Ak. Wiss.} (1916), p.688 (and  1918, p.154).}.

\par By proper normalization of the arbitrary function $\lambda$
in (\ref{eq:Wtwo}) one may impose the condition $\phi'=0$ upon the
$\phi_{i}$, thus giving Maxwell's equations a form often used by
H. A. Lorentz:

\begin{equation} \label{eq:Wnine}
   \Box \phi_{i} = s_{i}, \qquad  \partial \phi^{i} / \partial x_{i}=0.
\end{equation}

In the same manner one can choose the $\xi_{i}$ in
(\ref{eq:Weight}) so that $\gamma_{ik}=h_{ik}-\frac{1}{2}h \cdot
\delta_{ik}$ satisfies the equations

\begin{equation} \label{eq:Wten}
   \partial \gamma_{i}^{k} / \partial x_{k} = 0 \quad \textrm{and}
\end{equation}

\begin{equation} \label{eq:Weleven}
   \Box \gamma_{ik} = T_{ik}.
\end{equation}

\par In one important respect gauge invariance works differently
for electromagnetic and gravitational fields: If one splits the
tensor of derivatives $\phi_{k,i}=\partial \phi_{k}/ \partial
x_{i}$ into a skew and a symmetric part,

\begin{equation} \nonumber
   \phi_{k,i}= \frac{1}{2} (\phi_{k,i} - \phi_{i,k}) + \frac{1}{2} (\phi_{k,i} +
   \phi_{i,k}),
\end{equation}

the first part is not affected by a gauge transformation whereas
the second can locally be transformed into zero. In the
gravitational case \emph{all} derivatives $\partial h_{ik}/
\partial x_{p}$ can locally be transformed into zero. Hence we may
construct, according to Faraday and Maxwell, an energy-momentum
tensor $L_{ik}$ of the electromagnetic field,

\begin{equation} \label{eq:Wtwelve}
  L_{i}^{k}= f_{ip} f^{pk} - \frac{1}{2} \delta_{i}^{k}(ff), \qquad
  (ff)= \frac{1}{2} f_{pq} f^{qp},
\end{equation}

depending quadratically on the gauge invariant field components

\begin{equation} \nonumber
  f_{ik} = \phi_{k,i} - \phi_{i,k}  ,
\end{equation}

but no tensor $G_{ik}$ depending quadratically on the derivatives
$\partial h_{ik} / \partial x_{p}$ exists, if gauge invariance is
required, other than the trivial $G_{ik} \equiv 0$.

\vskip0.50cm
\textbf{2. Particles as centers of force, and the charge vector
and energy-momentum tensor of a continuous cloud of substance}.
Conceiving a resting particle as a center of force, let us
determine the \emph{static centrally symmetric solutions} of our
homogeneous field equations (\ref{eq:Wone}) and (\ref{eq:Wfive})
$(s^{i}=0$, $T_{ik}=0)$. One easily verifies that \emph{in the
sense of equivalence} the most general such solution is given by
the equations

\begin{equation} \label{eq:Wthirteen}
  \phi_{0} = e/4 \pi r, \qquad \phi_{i} =0  \quad \textrm{for }i \neq 0;
\end{equation}

\begin{equation} \label{eq:Wfourteen}
  \gamma_{00} = m/4 \pi r, \qquad \gamma_{ik} =0  \quad \textrm{for } (i,k) \neq (0,0) ;
\end{equation}

$r$ being the distance from the center. As was to be hoped, it
involves but two constants, \emph{charge} $e$ and \emph{mass} $m$.
The center itself appears as a singularity in the field. Indeed
$\phi_{0}$ and the factor $\phi$ in $\phi x_{a}$ $[\alpha=1,2,3]$
must be functions of $r$ alone, and the relations

\begin{equation} \nonumber
  \triangle \phi_{0} = 0, \qquad \partial \phi_{a}/ \partial x_{a} =0
  \qquad \qquad \qquad \qquad [\alpha=1,2,3]
\end{equation}

implied in (\ref{eq:Wnine}) then yield

\begin{equation} \nonumber
  \phi_{0} = a/r, \qquad \phi = b/r^{3},\qquad
  \phi_{a} = -(\partial/\partial x_{a})(b/r).
\end{equation}

Substitution of $\phi_{a}-\partial \lambda/ \partial x_{a}$ for
$\phi_{a}$ with $\lambda=-b/r$ changes $\phi_{a}$ into zero. In
the same manner (\ref{eq:Wfourteen}) is obtained from the
equations (\ref{eq:Wten} \& \ref{eq:Weleven}).

\par A continuous cloud of ``charged dust'' can be characterized
by its velocity field $u^{i}$ ($u_{i}u^{i}=1$) and the rest
densities $\mu$, $\rho$ of mass and charge. It is well known that
its equations of motion and the differential conservation laws of
mass and charge result if one sets $s^{i}=\rho u^{i}$ in Maxwell's
equations and lets $T_{i}^{k}$ in (\ref{eq:Wseven}) consist of the
Faraday-Maxwell field part (\ref{eq:Wtwelve}) and the kinetic part
$\mu u_{i}u^{k}$:

\begin{equation} \nonumber
  \partial (\rho u^{i})/ \partial x_{i} = 0, \qquad
  \partial (\mu u^{i})/ \partial x_{i} = 0; \qquad
  \mu \textrm{d}u_{i} / \textrm{d}s = \rho \cdot f_{ip}u^{p}.
\end{equation}

Since the motion of the individual dust particle is determined by
$\textrm{d}x_{i} / \textrm{d}s = u^{i}$ we have written
$\textrm{d} / \textrm{d}s$ for $u^{k}\partial / \partial x_{k}$.
In this manner Faraday explained by his electromagnetic tensions
(flow of momentum) the fact that the \emph{active} charge which
generates an electric field is at the same time the \emph{passive}
charge on which a given field acts.  At its present stage our
theory $(L)$ accounts for the force which an electromagnetic field
exerts upon matter, \emph{but the gravitational field remains a
powerless shadow}. From the standpoint of Einstein's theory this
is at it should be, because \emph{the gravitational force arises
only when one continues the approximation beyond the linear
stage}. We pointed out above that no remedy for this defect may be
found in a gauge invariant gravitational energy-momentum tensor.
However, the theory $(L)$ explains why active gravity, represented
by the scalar factor $\mu$ in the kinetic term $\mu u_{i}u_{k}$ as
it appears in the right member $T_{ik}$ of the gravitational
equations (\ref{eq:Wfive}), is at the same time inertial mass:
this is simply another expression of the fact that the mechanical
equations (\ref{eq:Wseven}) are a consequence of those field
equation.
\par We have seen that even in empty space the field part of
energy and momentum must not be ignored, and thus a particle
should be described by the static centrally symmetric solution of
the equations

\begin{equation} \label{eq:Wfifteen}
  D_{i}\phi=0,  \qquad
  D_{ik}h-L_{ik}=0
\end{equation}

(of which the second set is no longer strictly linear!). Again we
find, after proper gauge normalization,

\begin{equation} \label{eq:Wsixteen}
  \phi_{0} = e/4\pi r, \qquad
  \phi_{1} = \phi_{2} = \phi_{3} = 0,
\end{equation}

and then

\begin{displaymath}
\left\{ \begin{array}{ll} \label{eq:Wseventeen}
\gamma_{00} = m/4\pi r -1/4(e/4\pi r)^{2}, & \gamma_{0a}=0,\\
\gamma_{\alpha \beta} = -(e/4\pi r)^{2} \cdot (x_{\alpha}
x_{\beta}/4r^{2}) & \qquad \qquad \qquad [\alpha, \beta=1,2,3].
\end{array} \right.
\end{displaymath}

\vskip0.40cm As before, two characteristic constants $e$ and $m$
appear. \emph{At distance much larger than the ``radius''}
$e^{2}/4\pi m$ \emph{of the particle the gravitational influence
of charge becomes negligible compared with that of mass}.
\vskip0.50cm
\par $[$The remaining sections of the paper by Weyl are entitled:
sect. \textbf{3}. \textbf{The singular case}; sect. \textbf{4}.
\textbf{Derivation of the mechanical laws without hypothesis about
the inner structure of particles}; sect. \textbf{5}. \textbf{Vague
suggestions about a future unification of gravitation and
electromagnetism}; sect. \textbf{6}. \textbf{A free paraphrase of
Birkhoff's recent linear laws of gravitation} (\textbf{B}).$]$

\vskip1.00cm

\begin{center}
$^{\star---------------------------\star}$
\end{center}

\end{document}